# Crustal Structure Across the Northern Region of the Islas Marías Archipelago

Luis Alfredo Madrigal[1†], Diana Núñez[1*†], Felipe de Jesús Escalona-Alcázar[1,2†] and Francisco Javier Núñez-Cornú[1†]

[1]Centro de Sismología y Volcanología de Occidente (SisVOc), CUCosta, Universidad de Guadalajara, Puerto Vallarta, Mexico,
[2]Unidad Académica de Ciencias de la Tierra, Universidad Autónoma de Zacatecas, Zacatecas, Mexico

The tectonic interaction between the Rivera and North American plates north of the Bahía de Banderas is poorly understood. The nature of the crust and where the subduction ends in the western part of the Islas Marias Archipelago are still controversial. Based on new geophysical data provided by the TsuJal project, we present the shallow and deep crustal structure of the Rivera–North American plate contact zone along two seismic transects, TS09b and RTSIM01b, and the bathymetry obtained across the northern region of María Madre Island. Detailed bathymetric analysis allowed mapping of a series of lineaments along the study region, with two main preferred tendencies (020–050° and 290–320°) associated with the evolution of the Pacific-Rivera rise and the transform faults of the Gulf of California, respectively. The shallow structure is characterized by five sedimentary basins without deformation, whose horizons are subparallel, suggesting that the sediment deposition occurred after the extension process ended. The deep structure corresponds to a transition between oceanic crust (Rivera Plate), with an average thickness of ~10 km to the Islas Marías Escarpment, and a thinned continental crust, whose thickness increases toward the continent until it reaches 28 km, with a dip angle of 7–10°. The absence of an accretionary prism suggests that the subduction process of the Rivera Plate beneath the North American Plate to the north of Islas Marías has ceased. In this study, we determined that the morphological expression of the northern limit of the Rivera Plate corresponds to the Islas Marías Escarpment.

Keywords: Rivera Plate, crustal structure, amphibious network, horst and graben array, Islas Marías Archipelago

## INTRODUCTION

During most of the Cenozoic, the tectonic setting of Western Mexico has been a complex system of convergent, divergent, and transform boundaries between the various active tectonic units. The oblique subduction of the Farallon Plate beneath the North American Plate (NAP) progressively ceased since the Oligocene time as the Pacific-Farallon ridge approached North America and the triple junction moved southward. Along the Baja California Peninsula, from Miocene to recent, the movement of the triple junction produced the fragmentation of the Farallon Plate and formation of short-lived oceanic microplates, as well as the capture of the Baja California Peninsula by the Pacific Plate. All these processes, together with ridge rotation and the cessation of ridge activity along some segments west of the Baja California Peninsula, promoted plate reorganization (e.g., Mammerickx and Klitgord, 1982; Lonsdale, 1991; Bohannon and Parsons, 1995; Ferrari, 1995; Fletcher et al., 2007; Sutherland et al., 2012). After 12 Ma, the Pacific-Guadalupe ridge in the southern Baja California Peninsula broke up and rotated clockwise and oriented the rise to the NE (Mammerickx and

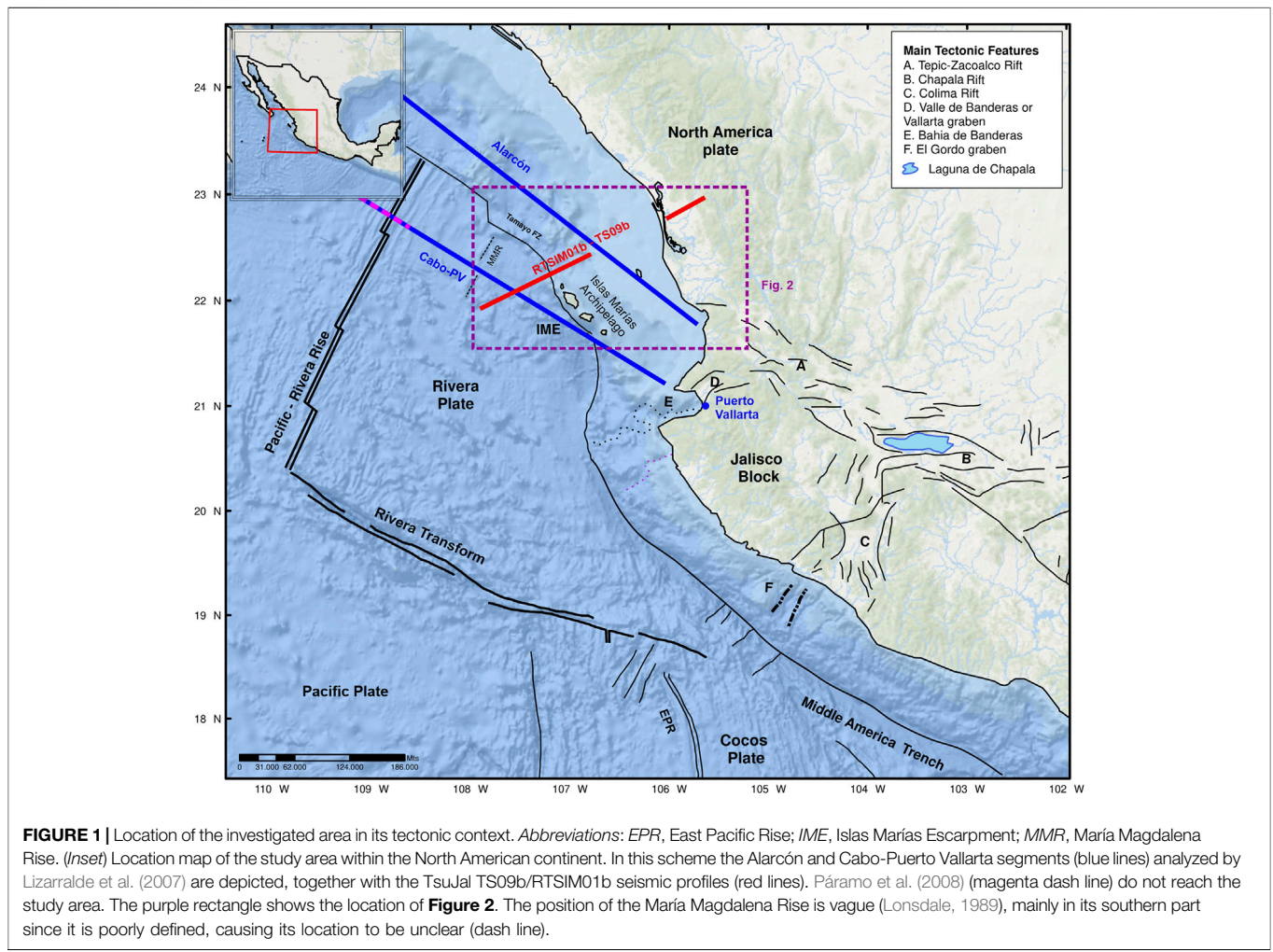

**FIGURE 1** | Location of the investigated area in its tectonic context. *Abbreviations*: EPR, East Pacific Rise; IME, Islas Marías Escarpment; MMR, María Magdalena Rise. (*Inset*) Location map of the study area within the North American continent. In this scheme the Alarcón and Cabo-Puerto Vallarta segments (blue lines) analyzed by Lizarralde et al. (2007) are depicted, together with the TsuJal TS09b/RTSIM01b seismic profiles (red lines). Páramo et al. (2008) (magenta dash line) do not reach the study area. The purple rectangle shows the location of **Figure 2**. The position of the María Magdalena Rise is vague (Lonsdale, 1989), mainly in its southern part since it is poorly defined, causing its location to be unclear (dash line).

Klitgord, 1982; Stock and Hodges, 1989). In this new plate configuration, the Pacific-Rivera and Pacific-Cocos ridges were formed (Mammerickx and Klitgord, 1982; Stock and Hodges, 1989; Lonsdale, 1991; Nicholson et al., 1994; Stock and Lee, 1994). The northern boundary of the Rivera Plate is defined by the Tamayo fracture zone (**Figure 1**), but the boundary extending from the Islas Marías Archipelago throughout Puerto Vallarta is less clear. Along this region, a lack of lithological and structural data, sedimentary cover, scarce outcrops on the islands, and issues with accessibility make it a challenge to get the necessary data to define this boundary precisely. Although several tectonic, structural, and marine geophysical studies were carried out, the continental crustal architecture and oceanic crust transition are not yet fully understood. Most of the geological and geophysical surveys are only in a few, limited places on islands, along seismic profiles, in mainland Mexico and the Baja California Peninsula (Fletcher et al., 2007; Housh et al., 2010; Sutherland et al., 2012; Ferrari et al., 2013; Pompa-Mera et al., 2013; Balestrieri et al., 2017). No study on the scale required to precisely define the plate boundary had yet been undertaken.

The Rivera Plate (RP), in terms of tectonic boundaries and motion, is problematic; it is an independent oceanic microplate (Atwater, 1970) of ~100,000 km² (DeMets and Stein, 1990) with distinct kinematics with respect to the NAP and Cocos Plate (Eissler and McNally, 1984; Bandy and Yan, 1989; DeMets and Stein, 1990). The RP is located to the west of Mexico; the northern limit is the Tamayo fracture zone. The eastern limit is the Middle America Trench subduction zone which truncates at the Tamayo fracture zone (e.g., Stoiber and Carr, 1973; Dean and Drake, 1978; Nixon, 1982; Bevis and Isacks, 1984; Ponce et al., 1992). The western limit of the RP is the Pacific-Rivera Rise, as part of the East Pacific Rise. The junction between the Pacific-Rivera Rise and the Rivera Transform is located 165 km west of the Middle America Trench (Bourgois and Michaud, 1991). The Rivera–Cocos Plate boundary runs from El Gordo graben and is continuous at depth along a line located east of the Colima Rift (Bandy et al., 1995). The crustal between RP and NAP at the northern part of the Islas Marías Archipelago is not clearly defined due to the low rate of recorded earthquake activity and the absence of bathymetric features of subduction in the zone (**Figure 1**).

To increase the structural understanding of the Islas Marías Archipelago and surroundings, a marine geophysical survey of the TsuJal project was performed (Núñez-Cornú et al., 2016).

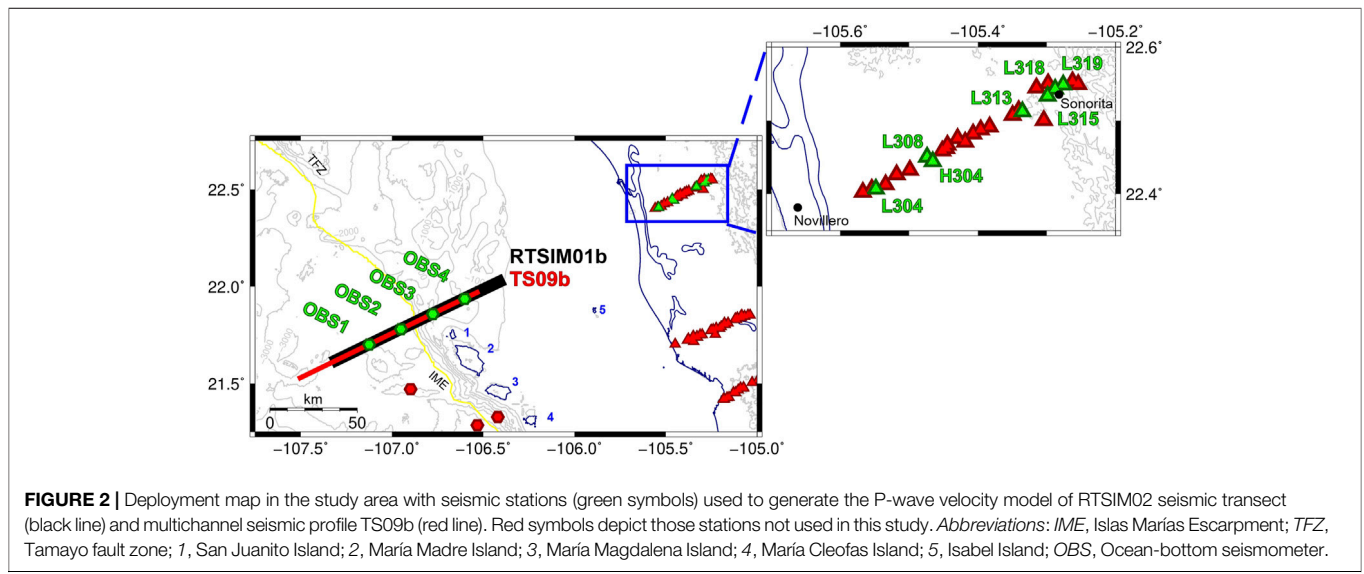

**FIGURE 2** | Deployment map in the study area with seismic stations (green symbols) used to generate the P-wave velocity model of RTSIM02 seismic transect (black line) and multichannel seismic profile TS09b (red line). Red symbols depict those stations not used in this study. *Abbreviations*: IME, Islas Marías Escarpment; TFZ, Tamayo fault zone; 1, San Juanito Island; 2, María Madre Island; 3, María Magdalena Island; 4, María Cleofas Island; 5, Isabel Island; OBS, Ocean-bottom seismometer.

**TABLE 1** | Seismic source parameters used during the RTSIM01b wide-angle (WAS) and the TS09b multichannel seismic (MCS) data acquisition.

| Seismic source parameters | WAS | MCS |
| --- | --- | --- |
| Source controller | Big Shot® | Big Shot® |
| Source type | Bolt® G.Guns 1500LL | Bolt® G.Guns 1500LL |
| Air pressure | 2000 psi | 2000 psi |
| Volume | 6,800 in$^3$ | 3,540 in$^3$ |
| Compressors | 4 x Hamworthy® 4$^{TH}$ 565 W100 | 4 x Hamworthy® 4$^{TH}$ 565 W100 |
| Number of air guns and strings | 11 air guns in 5 strings | 12 air guns in 4 strings (3 air gun/string) |
| Synchronization | ±0.1 ms | ±0.1 ms |
| Deployment depth | 15 m | 8 m |
| Trigger interval | 120 s | 50 m |

During the TsuJal project, several geophysical surveys were carried out combining sea–land studies with investigations involving the application of various geophysical techniques to characterize the surficial and crustal structure in the contact zone between RP and NAP. This article analyzes the results obtained from two seismic profiles, the RTSIM01b wide-angle seismic transect of 240 km and the lengthy TS09b multichannel seismic profile of 115 km, along with bathymetric data for this region. Both seismic lines are located in the north of the Islas Marías with SW–NE orientation (**Figure 2**) and are perpendicular to the contact between RP and NAP.

## DATA AND METHODOLOGIES

During the active part of the TsuJal project, the British research vessel *RRS James Cook* collaborated in acquiring multidisciplinary data (multichannel, wide-angle seismic, multibeam bathymetry and gravity and magnetism) in the western coast of Jalisco and Nayarit states. Moreover, this vessel deployed and collected the ocean-bottom seismometers (OBSs) and provided the seismic sources for the seismic experiment (**Table 1**).

### Bathymetric Data

The JC098 cruise provided the bathymetric and multichannel seismic data (MCS) analyzed in this work measured in the northern region of the María Madre Island, perpendicular to the coastline (**Figure 2**). Two multibeam echosounder systems (Kongsberg EM120 and EM710) acquired the bathymetric data used in this study. We also included the bathythermograph (XBT) probes and sound velocity profiles in the water column obtained daily during the data processing stage.

The bathymetric data recovered in the northern area of the TsuJal project across the RTSIM01b and TS09b seismic profiles were processed using CARIS HIPS and SIPS (Teledyne) software. We used sound speed and tide corrections provided by the Centro de Investigación Científica y de Educación Superior de Ensenada (CICESE) to produce vertical and horizontal data and georeferenced data, including calculating the total propagated uncertainty for each sounding. Finally, we obtained regular grid and variable resolution surfaces by applying various filters and editors to generate the final bathymetric surface with an 80 × 80 m resolution grid. This surface was interpolated and depicted using System for Automated Geoscientific Analyses (SAGA) and Geographic Information Systems (GIS) (Conrad et al., 2015).

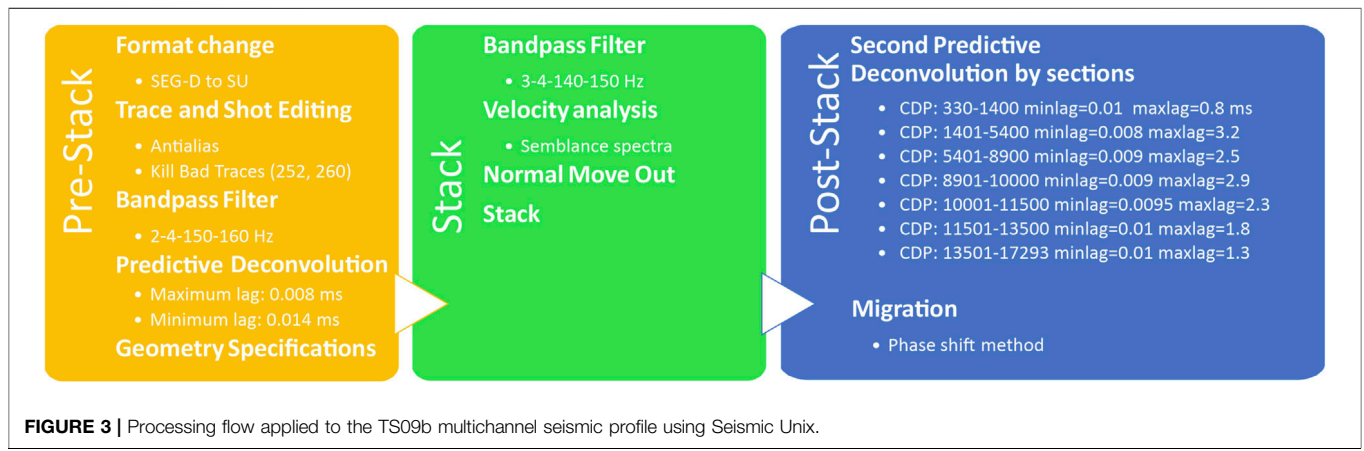

FIGURE 3 | Processing flow applied to the TS09b multichannel seismic profile using Seismic Unix.

## Multichannel Seismic Data

The MCS data were acquired by using a SASS Multichannel Sentinel Sercel® streamer of 5.85 km length (468 active channels, separated 12.5 m) deployed at 10 m depth. The common depth point (CDP) distance is 6.25 m, providing a CDP nominal fold of 58–59 traces. These data were recorded initially in SEG-D format and sampled at 1 ms. The technical parameters of the seismic source used in this study are shown in **Table 1**. The TS09b seismic line consisted of 2,305 shots with a total length of 115 km approximately.

**Figure 3** shows the main steps of the processing stage, which was carried out by Seismic Unix software (Cohen and Stockwell, 2013). We carried out a traditional processing methodology to increase the horizontal and vertical resolution to obtain the best possible seismic image of the TS09b seismic profile. The sequence shown in **Figure 3** includes the following steps:

1. Pre-stacked signal calculations (eliminate aliasing, eliminate incorrect traces, and filtering)
2. Spherical corrections and predictive deconvolution for improving the resolution in time
3. Velocity analysis by semblance method every 100 CDP
4. Correction of normal move out
5. Stack to increase the signal-to-noise ratio
6. Phase shift migration with turning rays for increasing horizontal resolution and collapse diffractions, which relocate the reflectors in time.

## Wide-angle Seismic Data Acquisition and Seismic Phases

The wide-angle seismic data correspond to a SW–NE trending line (RTSIM01b) along the north of the María Madre Island, perpendicular to the coastline (**Figure 2**). The length of the shooting line was 110 km approximately. These shots were recorded on land 240 km from the first shot of the transect. The seismic source consisted of two air gun subarrays with a total capacity of 6,800 in$^3$, shooting every 120 s, and with a cruise velocity of five knots (**Table 1**). In addition, these shots were registered by an amphibious seismic network composed of four OBSs and 27 temporary seismic stations deployed from Novillero to Sonorita (Nayarit) (**Figure 2**). The OBSs were short-period model LC2000SP with L-28 three-component geophone sensors (4.5 Hz) and one HiTech HYI-90-U hydrophone (OBS1, OBS2, OBS3, and OBS4). The instruments used for the terrestrial network were single, vertical component TEXAN 125A (Reftek, Trimble), a TAURUS Digital Seismograph (Nanometrics), and WorldSensing Spidernano data loggers both with a short-period sensor 1-Hz LE-3D/lite (Lennartz) and a CMG-6TD (Güralp Systems). Of the 27 seismic stations deployed along this line, we have selected those whose signal-to-noise ratio was low, resulting in seven high-quality stations (L304, L308, H305, L313, L315, L318, and L319) (**Figure 2**).

The data processing included band-pass filtering and navigation data. Instrumental drift corrections, zero-phase band-pass filter (4–10 Hz), and travel time corrections were also applied (Núñez et al., 2016). Furthermore, topography and bathymetric data were included for P-wave phase determination (**Figures 4**, **5**), which consisted of correlating reflected and refracted phases observed at the different crust and uppermost mantle discontinuities. We calculated the apparent velocities from P-wave refracted phases used for initial velocity and depth modeling. We identified five refracted phases [three within the sediments ($P_{S1}$, $P_{S2}$, and $P_{S3}$), one within the crust ($P_I$), and one within the uppermost mantle ($P_n$)] and four reflected phases [one intermediate-lower crust discontinuity ($P_{LC}P$), one crust-mantle boundary reflection ($P_MP$), and two reflections in the first layers of the upper mantle ($P_{M1}$ and $P_{M2}$)].

The sedimentary cover across the RTSIM01b transect was sampled by the refracted phases mainly observed in OBS seismic record sections (**Figure 4**). The $P_{S1}$, $P_{S2}$, and $P_{S3}$ phases were identified from 3 to 15 km of the source–receiver offset distance for the shallowest refracted phases, while between 6 and 30 km for the third one, in most OBS sections. The average apparent velocities calculated were 2.8, 4.1, and 5.1 km/s, respectively. The next phase, $P_I$, is observed in the offset interval 10–30 km for the marine record sections, and it was identified from 96 to 111 km in the station closest to the coast (L304) with 5.8 km/s. The $P_{LC}P$ seismic phase is correlated between 10 and 46 km offset distance for OBS1 and 30–45 km for OBS2 and OBS3 (**Figure 4**). All of the seismic sections exhibit a secondary arrival, $P_MP$, indicating an abrupt discontinuity between the crust and the upper mantle. In both sections recorded by marine and land stations, we identified the Pn phase with an average apparent velocity of 8.1 km/s offshore and 8.5 km/s for the onshore region. The uppermost mantle discontinuities are primarily identified in the temporary land stations in the offset interval 135–170 km for $P_{M1}$ and 170–220 km of source–receiver offset distance for $P_{M2}$ (**Figure 5**).

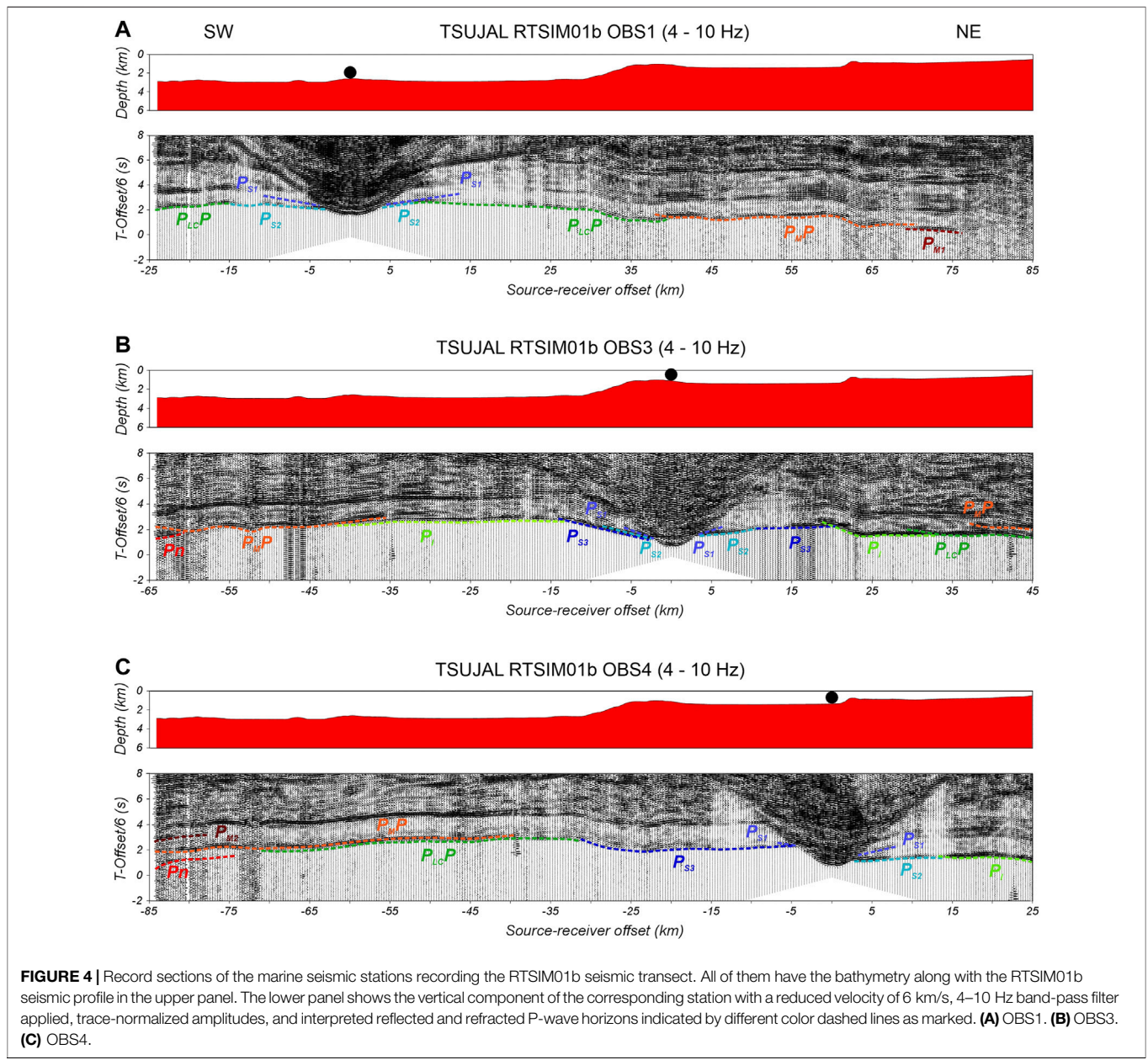

**FIGURE 4** | Record sections of the marine seismic stations recording the RTSIM01b seismic transect. All of them have the bathymetry along with the RTSIM01b seismic profile in the upper panel. The lower panel shows the vertical component of the corresponding station with a reduced velocity of 6 km/s, 4–10 Hz band-pass filter applied, trace-normalized amplitudes, and interpreted reflected and refracted P-wave horizons indicated by different color dashed lines as marked. **(A)** OBS1. **(B)** OBS3. **(C)** OBS4.

A total of 1,617 arrivals were manually picked, defining the seismic phases identified throughout changes in amplitude or frequency content with an average estimated picking error of 108 ms. The best 2D velocity and interface structure model that fits the previous WAS data was obtained using the Zelt and Smith (1992) software package, applying forward modeling, travel time inversion, and synthetic seismograms.

# RESULTS

## Multichannel Seismic and Bathymetric Data

Along the JC098 cruise track, we obtained the bathymetric data around the Islas Marías Archipelago (**Figure 6**). The detailed bathymetry is shown in **Figure 6A**, whereas a 3D perspective is shown in **Figure 6B**. Based on the alignment of the submarine relief, we interpreted 88 structural lineaments and calculated a rose diagram from their azimuths. The detected lineaments could be either faults or fractures, with the former no longer being active, and provide the preferred structural trends of the region (**Figure 6C**). Two main tendencies were obtained: 1) between 020 and 050° and 2) between 290 and 320°. Both tendencies are spatially well defined. The first one is located west and northwest of the archipelago, probably related to the RP oceanic crust. The second trend is placed at the southwestern and west sides of the islands within the transitional or continental crust of the NAP. In the west area of María Madre and María Magdalena islands, the structural lineaments have an ENE–WSW trend, which is oblique to the main trend from the adjacent areas.

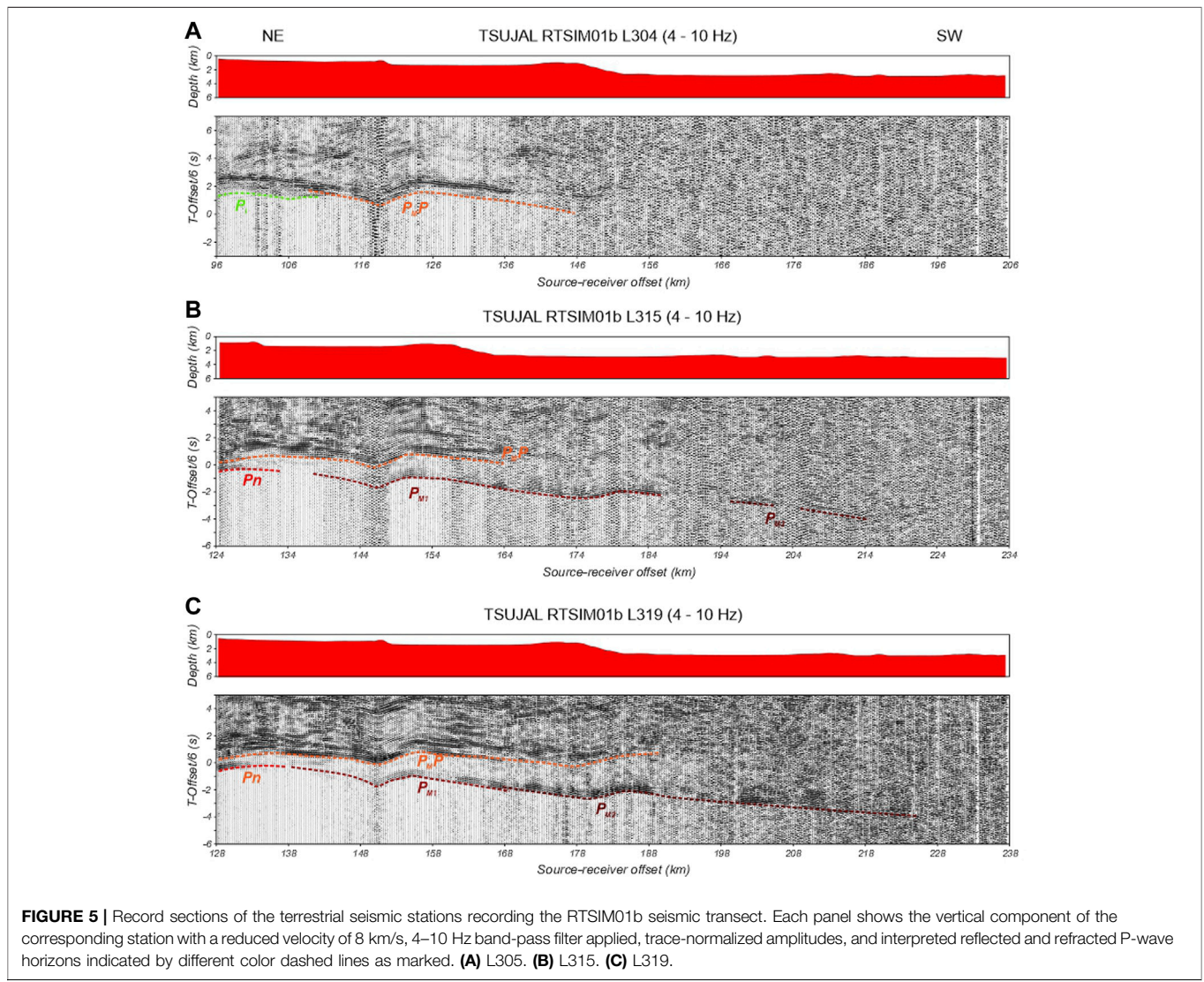

FIGURE 5 | Record sections of the terrestrial seismic stations recording the RTSIM01b seismic transect. Each panel shows the vertical component of the corresponding station with a reduced velocity of 8 km/s, 4–10 Hz band-pass filter applied, trace-normalized amplitudes, and interpreted reflected and refracted P-wave horizons indicated by different color dashed lines as marked. **(A)** L305. **(B)** L315. **(C)** L319.

The study of shallow structures to the north of Islas Marías Archipelago included the seismic transect TS09b (**Figure 7**), where we were able to identify two main seismic facies: the acoustic basement and different sedimentary packages distributed in five basins. Three of them are on the RP (Rivera, North Rivera, and Islas Marías basins) and two on the NAP (East Nayarit and San Blas troughs). From SW to NE, we found part of the Rivera Basin located between marks 0 and 15 km. It is infilled by up to 0.3 s of two-way travel time (twtt) of sediments. Toward the northeast, the North Rivera Basin is observed between 27 and 40 km with a thickness of 0.6 s approximately (**Figure 7**). In this basin, a structure rises from the acoustic reflector, which could be a volcano, as Dañobeitia et al. (2016) suggested, or a horst since it is fault-bounded. Moreover, the sedimentary horizons are subhorizontal on both sides of the horst.

The Islas Marías Basin corresponds to the largest basin identified along our MCS profile, extending between 45 and 77 km infilled by up to 1 s (twtt) of sediments. This basin is limited to the SW by the María Range and to the NE by the Islas Marías Escarpment (**Figures 6**, **7**). The sedimentary horizons are subparallel, and we identified a fault, not reported in previous studies, located to the SE of the Tres Marías Fault, crosscutting the lower part of the sedimentary infill. Located to the northeast side of the Islas Marías Escarpment, in the southernmost part of the East Nayarit Trough, we find the deepest basin between 95 and 110 km (15 km) infilled by up to 1.5 s sediments, bounded by the Oriental Nayarit Fault on the eastern edge of the trough. The sedimentary horizons have a splay array toward the Oriental Nayarit Fault. This feature is not observed in any other basins, suggesting that the Oriental Nayarit Fault could be active. The northeasternmost basin corresponds to the San Blas Trough with a 3-km width and 0.5 s (twtt) of thickness whose sedimentary horizons are subhorizonal.

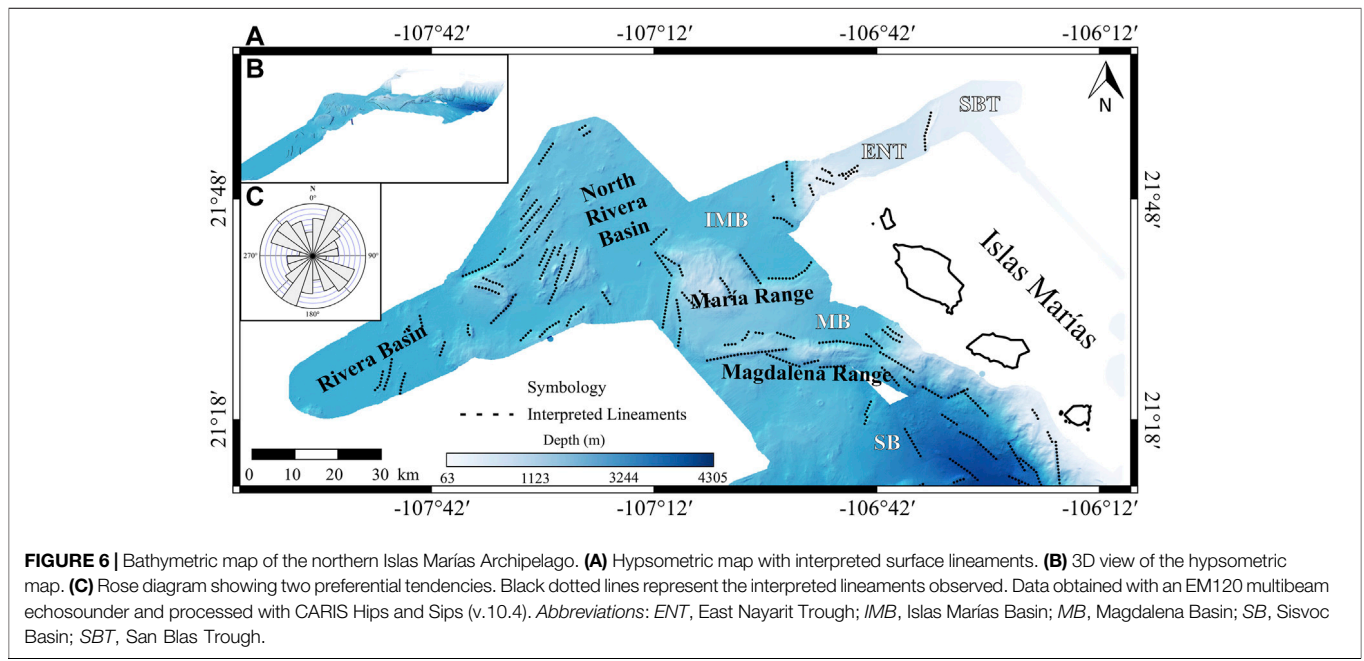

**FIGURE 6** | Bathymetric map of the northern Islas Marías Archipelago. **(A)** Hypsometric map with interpreted surface lineaments. **(B)** 3D view of the hypsometric map. **(C)** Rose diagram showing two preferential tendencies. Black dotted lines represent the interpreted lineaments observed. Data obtained with an EM120 multibeam echosounder and processed with CARIS Hips and Sips (v.10.4). *Abbreviations*: *ENT*, East Nayarit Trough; *IMB*, Islas Marías Basin; *MB*, Magdalena Basin; *SB*, Sisvoc Basin; *SBT*, San Blas Trough.

The basement and the acoustic basement along the profile showed extensional deformation, indicating a horst and graben array (**Figure 7**). In general, the sediment horizons are not deformed within the basins, suggesting that the extension finished before the sedimentation started. However, the splay array adjacent to the Oriental Nayarit Fault could be related to some activity along this fault. Along this transect, neither have we detected the presence of an accretion prism.

## WAS Data

The final P-wave velocity model corresponding to the wide-angle seismic profile RTSIM01b is an offshore–onshore transect of 240 km length, which characterizes the northern region of the Islas Marías Archipelago tectonically (**Figure 8**). The profile's origin was located at 24 km between the OBS1 and the shot situated farther to the southwest. We divide our model according to P-wave velocities in the upper crust and sedimentary cover, the crust (middle and lower), and the upper mantle.

The sedimentary cover along the RTSIM01b profile is characterized by three basins with different depths and velocities. From SW to NE, we found the North Rivera Basin is located from 25 to 53 km from the model origin with an average thickness of 1 km with a maximum value of ~2 km around OBS2 and a velocity range of 2.5–4.2 km/s. From 65 to 85 km of model distance (**Figure 8**), the Islas Marías Basin corresponds to the deepest basin across the transect with ~3 km of thickness, and for the upper crust or sediment layer, the average velocity is 3.6 km/s, and velocities from top to bottom vary between 2.5 and 4.6 km/s. The northeasternmost basin, the East Nayarit Trough, has a P-wave velocity interval of 2.8–3.8 km/s and 1.3 km thick. The middle crust below the sedimentary cover has a vertical velocity gradient of 5.3–5.8 km/s, with thickness increasing to the northeast direction, reaching 10 km depth in the continental region. Two layers comprise the lower crust, where the thickness of the oceanic crust is ~6 km, increasing toward the northeast up to 10 km thick, and P-wave velocity changes for the upper and lower layers are 6.3–6.5 km/s and 6.7–6.9 km/s, respectively. The Moho depth reaches 10 km depth in the Rivera Plate region, thickening up to 28 km in the continental part of the model with a velocity contrast of 6.9 km/s to 7.9 km/s (**Figure 8**). In the upper mantle, we characterized two seismic layers with increasing velocity at a depth from 8.1–8.4 km/s down to 40 km.

After adjusting travel times, we controlled by amplitudes using synthetic seismograms to get our final P-wave velocity model. This model reproduces 1,592 of 1,617 (98%) of observed travel times throughout the entire length of the profile (240 km). We determined the arrival-time fit quality ($\chi_N^2$) for each interpreted phase with the following values for $P_{S1}$ (0.4), $P_{S2}$ (0.3), $P_{S3}$ (1.2), $P_I$ (1.3), $P_{LC}P$ (0.7), $P_{LC}$ (0.9), $P_MP$ (2.4), and $P_n$ (0.9), and reflected P-phases observed in the mantle $P_{M1}$ (1.1) and $P_{M2}$ (0.8). Our final model is not far from the ideal case ($\chi_N^2 = 1$), producing a $\chi_N^2$ of 1.6.

## DISCUSSION

The complex architecture of Rivera and North American plate interaction has been studied in the northern region of Islas Marías Archipelago. Previous bathymetric studies have been reported in this area, but few have been analyzed together with multichannel seismic profiles (Lizarralde et al., 2007 (**Figure 1**); Sutherland et al., 2012; Dañobeitia et al., 2016; Carrillo-de la Cruz et al., 2019). Most of the studies tried to establish the nature of the crust of the Islas Marías, which is still under debate. Lonsdale (1989) considered the Islas Marías to be a block of continental crust, while Schaaf et al. (2015a, 2015b) suggested that in María Magdalena Island, the lithologic assemblage includes oceanic crust. Additionally, Lizarralde et al. (2007) suggested a transitional crust, while Carrillo-de la Cruz et al. (2019) suggested a thinned continental crust. Our structural interpretation of the bathymetric features (**Figure 6**) indicates two preferred orthogonal tendencies, whose location and orientation indicate deformation in the oceanic and thinned continental crust. The structural trend in the thinned continental crust (**Figure 6C**) ranges from 290 to 320°, which is subparallel to the 305° azimuth of the transform faults within the Gulf of California (Lonsdale, 1989). The orthogonal structures are associated with the Pacific-Rivera rise evolution (Lonsdale, 1989).

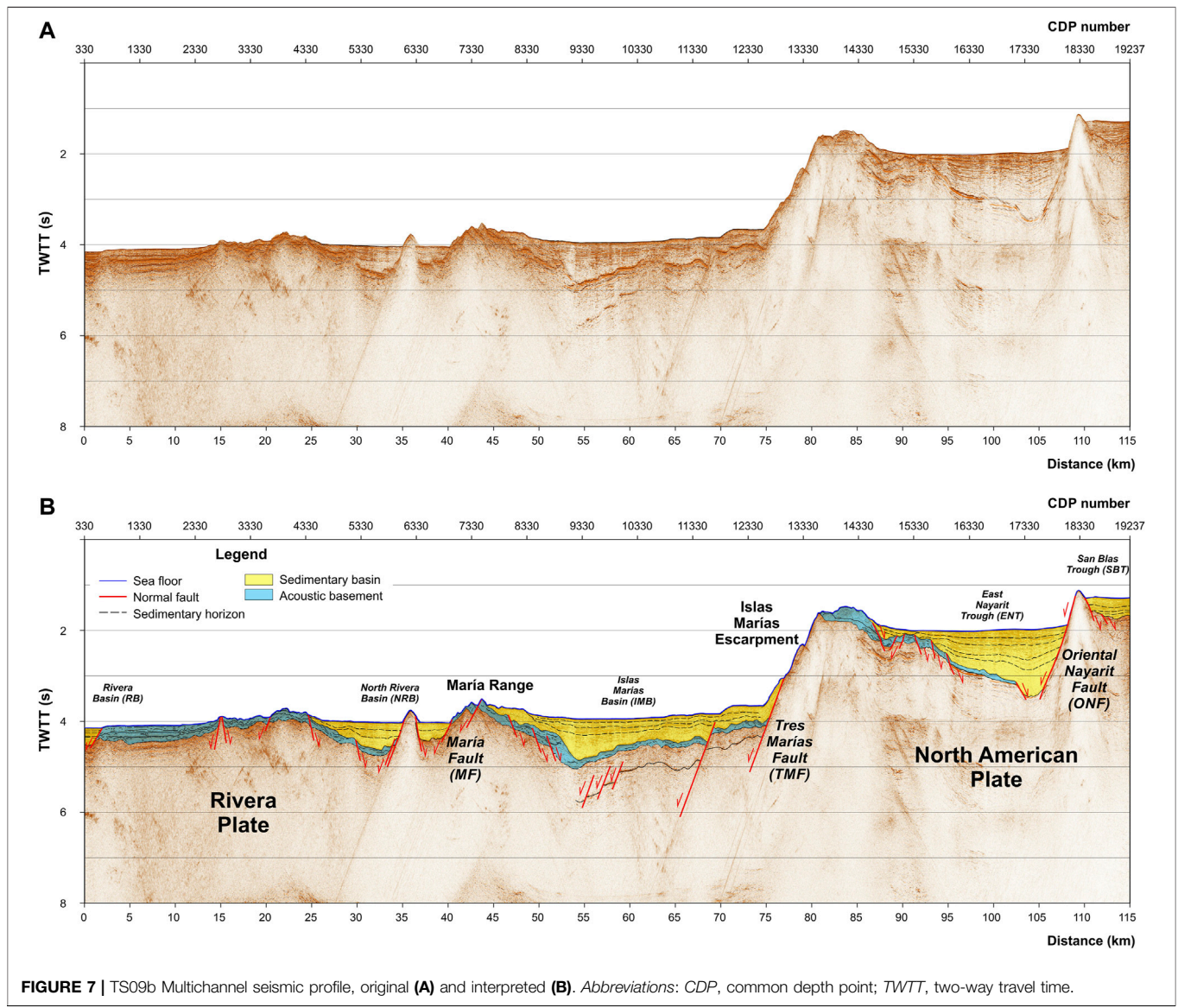

FIGURE 7 | TS09b Multichannel seismic profile, original (A) and interpreted (B). *Abbreviations*: CDP, common depth point; TWTT, two-way travel time.

The seismic profiles (**Figure 2**) show a P-wave velocity distribution of <6 km/s that corresponds to the continental crust that thins (**Figure 8A**) and extends west and NW of the Islas Marías Archipelago (Acosta-Hernández, 2017; Dañobeitia et al., 2016; Carrillo-de la Cruz, 2017; Madrigal-Ávalos, 2018; Carrillo-de la Cruz et al., 2019). The subduction of the RP beneath of NAP at the southern contact is clearly defined with a dip angle of 12–14° (Núñez et al., 2019), with the Middle America Trench acting as the morphological expression of the contact between them. To the NW of the Middle America trench, the only bathymetric expression of the contact between RP and NAP is the Islas Marías Escarpment. The Islas Marías Escarpment is a normal fault reported in its southern part by Carrillo-de la Cruz et al. (2019). Some authors establish the location of the active subduction west of our profile at ca. 15 Ma (Lonsdale, 1989) and ca. 12 Ma (Fletcher et al., 2007; Sutherland et al., 2012; Ferrari et al., 2013; Duque-Trujillo et al., 2014). The dip angle obtained from our P-wave velocity model is 7–10° (**Figure 8**), according to the values reported by Dañobeitia et al. (1997).

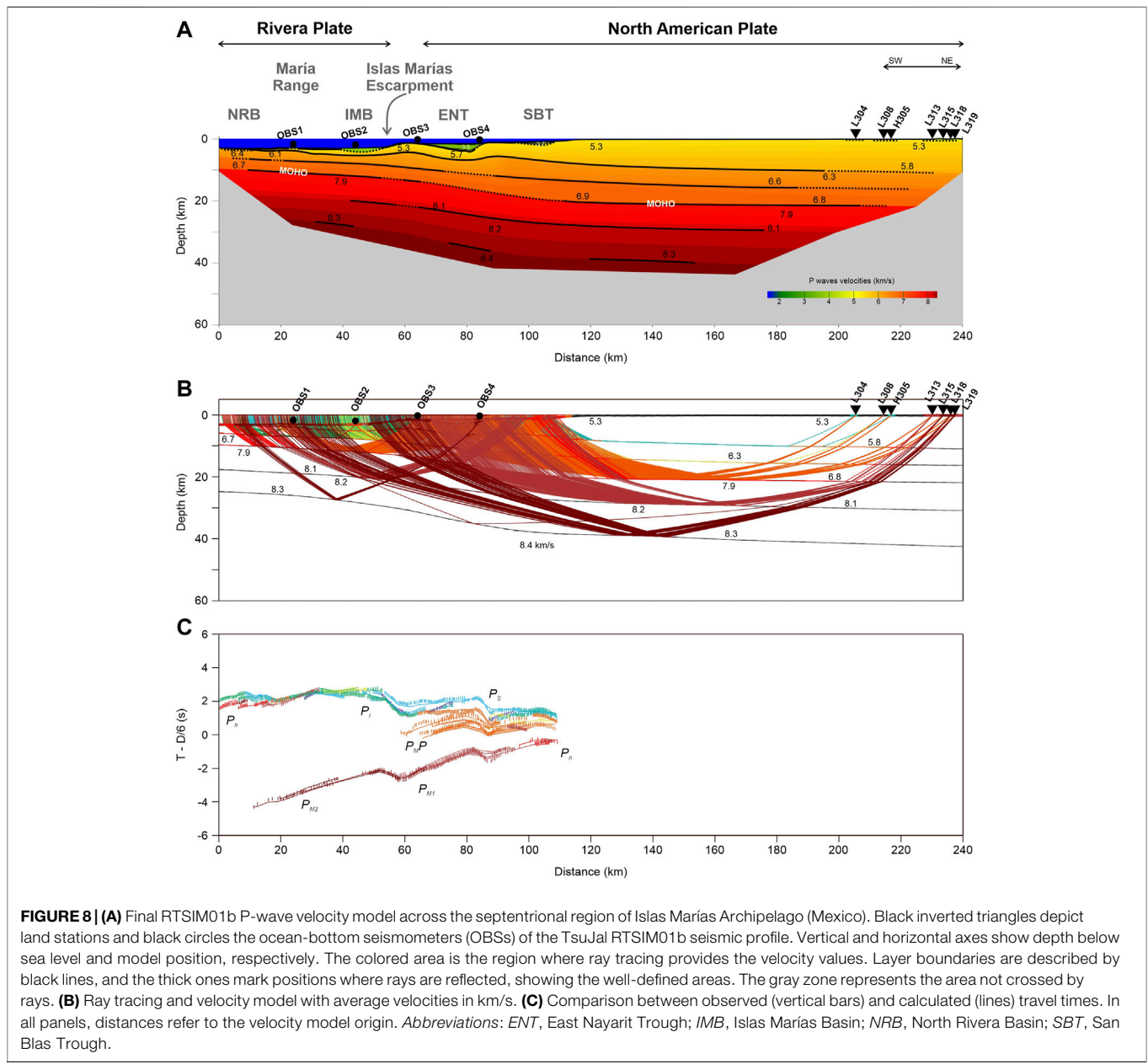

**FIGURE 8 | (A)** Final RTSIM01b P-wave velocity model across the septentrional region of Islas Marías Archipelago (Mexico). Black inverted triangles depict land stations and black circles the ocean-bottom seismometers (OBSs) of the TsuJal RTSIM01b seismic profile. Vertical and horizontal axes show depth below sea level and model position, respectively. The colored area is the region where ray tracing provides the velocity values. Layer boundaries are described by black lines, and the thick ones mark positions where rays are reflected, showing the well-defined areas. The gray zone represents the area not crossed by rays. **(B)** Ray tracing and velocity model with average velocities in km/s. **(C)** Comparison between observed (vertical bars) and calculated (lines) travel times. In all panels, distances refer to the velocity model origin. *Abbreviations*: ENT, East Nayarit Trough; IMB, Islas Marías Basin; NRB, North Rivera Basin; SBT, San Blas Trough.

Along the TS09b profile (**Figure 7**), the identified structures are normal faults in a horst and graben array. Most of these faults are currently inactive, as suggested by the subhorizontal sedimentary horizons within the basins and troughs, and the lack of seismicity. Moreover, the normal faults truncate at the surface of the acoustic basement and do not extend into the sedimentary deposits of the basins (**Figure 7B**). The only structure that could have some extant seismic activity is the Oriental Nayarit Fault since the sediments have a splay array that becomes horizontal at the top. No other fault seems to be active along the TS09b profile.

Furthermore, our study clarifies the nature of two structures previously not adequately located in the study region. One is the fault-bounded María Range (Escalona-Alcázar et al.

submitted), which was previously mistakenly reported as the María Magdalena Rise (Dañobeitia et al., 2016). However, according to Lonsdale (1989), the María Magdalena Rise is situated to the west of the TS09b profile (**Figure 1**). The southern segment of the María Magdalena Rise (**Figure 1**) is not well defined in extension and length (Lonsdale, 1989), so this segment would intersect the TS09b/RTSIM01b profiles at ~15 km from its beginning. At this length, there is a tiny horst-like structure (~0.5 km wide) (**Figure 7B**), and, 7.5 km to the NE, another horst of ~ 4 km wide is located. It is not clear if one of them could be related to the southern segment of the María Magdalena Rise. The first ~ 30 km of the TS09b/RTSIM01b profiles have a pop-up–like shape, with a horst and graben array with no sediments in between (**Figures 7, 8**). The bathymetric heights at 15 and 35 km from the profile starting point were interpreted as a volcano (Dañobeitia et al., 2016). Nonetheless, they could be horst since the normal faults go from the surface to below the acoustic basement (**Figure 7A**); additionally, the P-wave velocities shown in **Figure 8A** are of < 6 km/s, typical of the continental crust.

Between Los Cabos and Puerto Vallarta, the faults were mainly oriented NE–SW due to extension at the mouth of the Gulf of California after ca. 6 Ma (Lonsdale, 1989; Stock and Hodges, 1989; Sutherland et al., 2012; Abera et al., 2016). At the same time, the extension expanded to the east, thinning the continental crust. In this scenario, the role of the María Magdalena Rise is unclear, but its orientation (azimuth 025°; Lonsdale, 1989) parallel to the Tamayo and Nayarit troughs, both adjacent to the mainland Mexico, as well as the structural trend of the María Madre Island (Escalona-Alcázar et al., 2014), suggests a wide area of extension. Around the TS09b profile, this extension could have ceased at ca. 3.5 Ma, when the María Magdalena Rise ended its activity (Lonsdale, 1989); then the sedimentary fill started.

## CONCLUSION

The analysis and interpretation of the study carried out in the northern region of the Islas Marías Archipelago provide new information about the structure and tectonics of the region, where it is possible to establish that Rivera Plate subduction under the North American Plate has likely ceased or never took place at this location. Nevertheless, we determined that the morphological expression of the northern limit of the Rivera Plate is the Islas Marías Escarpment.

The average crustal thickness for the Rivera Plate is ~10 km up to the Islas Marías Escarpment, estimating a depth of Moho deeper than 13 km in the collision zone between both tectonic plates. The crust of the North American Plate thickens from the Islas Marías Escarpment to the NE, up to reach 28 km.

From the MCS seismic image, it has been possible to characterize five sedimentary basins without deformation associated with compressional movements, where the absence of an accretionary prism is also relevant, demonstrating there is no active subduction process in this region. Sedimentary horizons in all basins are subhorizontal, suggesting that they were deposited after extension in the area ended during the late Pliocene. Only the Oriental Nayarit Fault could possibly support some seismic activity.

## DATA AVAILABILITY STATEMENT

## AUTHOR CONTRIBUTIONS

LM: data processing, investigation, and original draft writing. DN: conceptualization, methodology, investigation, data processing, data analysis, conclusions, and original draft writing. FE-A: data analysis, methodology, investigation, original draft writing, and conclusions. FN-C: conceptualization, funding acquisition, methodology, investigation, review, and editing.

## FUNDING

This research was mainly funded by Consejo Nacional de Ciencia y Tecnología (CONACYT)—FOMIXJal (2012-08-189963) (Mexico) and CGL (2011–29,474-C02–01) DGI Plan Nacional I + D + i (Spain) (TsuJal Project). LM was financially supported by a master fellowship from CONACyT with code 422412 and CVU704296.

## ACKNOWLEDGMENTS

We gratefully thank Wendy McCausland for her valuable comments and observations, including the English grammar revision, Juan Luis Carrillo-de la Cruz for his help with bathymetric figure and the suggestions of two reviewers that help to improve this article. The authors also truly appreciate the colaboration during the TsuJal project of NOC Cruise JC098, RRS James Cook (United Kingdom); COIP/COPO/UNAM J-GAP2013 Cruise (BO El Puma); Unidad de Tecnología Marina (Spain); Secretaría de Marina (Mexico) ARM Holzinger; Secretaría de Defensa Nacional (Mexico); Unidad Municipal de Protección Civil y Bomberos (Jalisco State, Mexico); Unidad Municipal de Protección Civil y Bomberos (Puerto Vallarta, Mexico); Unidad Estatal de Protección Civil y Bomberos (Nayarit State, Mexico); Reserva de la Biosfera (Islas Marías) CONANP-SEMARNAT; Secretaría de Relaciones Exteriores (Mexico); and Órgano Desconcentrado de Prevención y Readaptación Social de la SEGOP. Some figures were generated using the Generic Mapping Tools (GMT-6; Wessel et al., 2019).

**Conflict of Interest:** The authors declare that the research was conducted in the absence of any commercial or financial relationships that could be construed as a potential conflict of interest.